\documentclass[final]{eptcs}
\providecommand{\event}{ICE 2018} 
\usepackage{breakurl}             
\usepackage{underscore}           

\usepackage{amsthm}
\usepackage{enumerate}
\usepackage{mathtools} 
\usepackage{amsmath}
\usepackage{amssymb}
\usepackage{eucal}
\usepackage[version=0.96]{pgf}
\usepackage{tikz}
\usetikzlibrary{arrows,shapes,snakes,automata,backgrounds,petri}
\usepackage{stmaryrd,epsfig,color,moreverb}
\newtheorem{definition}{Definition}
\newtheorem{theorem}{Theorem}
\newtheorem{example}{Example}
\newtheorem{proposition}{Proposition}

\newcommand{\trans}[1]{\ensuremath{\,[\/{#1}\/\rangle}\,}

\newcommand{\pre}[1]{\ensuremath{\!~^\bullet{#1}}}

\newcommand{\post}[1]{\ensuremath{{#1} {^\bullet}}}

\newcommand{\nat}{\ensuremath{\mathbb{N}}}

\newcommand{\eqclass}[2]{\ensuremath{(\!|#1|\!)_{#2}}}

\newcommand{\conf}[2][]{\mathcal{C}\mathit{onf}_{{#1}}({#2})}

\renewcommand{\epsilon}{\varepsilon}
\newcommand{\hidden}[1]{}

\newcommand{\setenum}[1]{\{#1\}}
\newcommand{\setcomp}[2]{\{{#1} \;\mid\; {#2}\}}

\newcommand{\card}[1]{\ensuremath{|#1|}}
\newcommand{\partition}{\ensuremath{\nu}}

\newcommand{\SafeNet}{\ensuremath{\mathbf{Safe}}}
\newcommand{\MultiClockNet}{\ensuremath{\mathbf{MCN}}}

\newcommand{\mcn}{{\textsl{mc}-net}}
\newcommand{\ind}[1]{\ensuremath{\upsilon(#1)}}
\newcommand{\vcd}{\textsc{vcd}}

\newcommand{\netdom}{\textsc{nd}}

\newcommand{\Spread}{\ensuremath{\mathbf{Spread}}}

\newcommand{\place}{\ensuremath{P}}
\newcommand{\dime}[1]{\ensuremath{\iota(#1)}}

\title{Toward a Uniform Approach to the Unfolding of Nets}

\author{Eric Fabre
\institute{INRIA Rennes - Bretagne Atlantique, France
\email{eric.fabre@inria.fr}
}
\and
G.~Michele Pinna
\institute{Universit\`a degli Studi di Cagliari, Italy 
\email{gmpinna@unica.it} 
}
}

\def\titlerunning{Toward a Uniform Approach to the Unfolding of Nets}
\def\authorrunning{E. Fabre \& G. M. Pinna}

\begin{document}
\maketitle

\maketitle

\begin{abstract}
In this paper we introduce the notion of \emph{spread} net. Spread nets are (safe) Petri nets equipped with vector clocks on places and with ticking functions on transitions, and are such that vector clocks are consistent with the ticking of transitions. Such nets generalize previous families of nets like unfoldings, merged processes and trellis processes, and can thus be used to represent runs of a net in a true concurrency semantics through an operation called the spreading of a net. By contrast with previous constructions, which may identify conflicts, spread nets allow loops in time.
\end{abstract}

\section{Introduction}
One of the most popular motto in Petri nets is that ``\emph{the semantics of a net is a net}'' (\cite{SR:SNN87}). 
Along this line of thought \emph{non sequential} processes have been proposed (\cite{GR:NSBPN83}), where the 
causal dependencies among the transitions of a net are faithfully represented.
To model all the possible non sequential executions of nets, the notion of \emph{unfolding} of a net has
been proposed in \cite{NPW:PNES} and further investigated in \cite{Win:ES} and \cite{Eng:BPPN}. The idea is
to represent conflicts as branching alternatives (whence the name of \emph{Branching Processes}, that are essentially 
unfoldings).

Unfoldings were introduced to represent the non sequential behaviors of (safe) Petri nets, but their main application
originated in the fact that they offered new techniques for the verification of concurrent systems: rather
than exploring the sequential behaviors of nets the use of partial orders allows one to have more 
compact representations
of these behaviors. Still, the data structure is in general infinite or too large. 
One of the first attempts to overcome this problem was to turn non sequential processes into an algebra, with a 
parallel composition and a suitable notion of \emph{concatenation} (\cite{DMM:ANCP89} and further investigated
in \cite{DMM:AANCP96}). An orthogonal approach has been the one pursued in \cite{McMillan92} where
the unfolding is \emph{cut} in a way that still allows one to infer all the information needed to represent 
all possible
computations of a safe net. 
Another way to address this problem is to define an equivalence on some behaviors of (safe) Petri nets, which
implies that the data structure adopted cannot be any longer the one devised for unfoldings or prefixes. 
The notion of \emph{unravel} net introduced in \cite{CP:Soap17} and \cite{CP:Lata17} goes in this 
direction requiring that each execution is a partial order, but the overall structure does not need to
be a partial order.
Overcoming the request that (at least locally) the behavior should be represented using a partial order
has led to the introduction of a \emph{reveal} relation playing the role of causality 
\cite{BCH:BonRR,HKS:crron}. There it is shown how to relate occurrence nets and reveal relations. 
Still the more compact data structure has its origin in the partial ordering representing the dependencies in the
net. 

In this paper we face the problem from another point of view. Rather than focussing on the properties the whole
net representing the behavior of another net has to enjoy, we enrich the net with informations that will
play a role analogous to those played by the properties the data structure has to fulfil.

We focus on systems (nets) that are composed of simpler subsystems: 
basically finite state automata. 
These automata \emph{synchronize} on common transitions. The resulting system gives us the basic ingredients we want to
elaborate on: causality, that coincides with \emph{time} in each subsystem, and conflicts, which are local to a component as well. 
Each synchronization among finite state automata determines the expansion of conflicts and causalities to all the components of the system.
When \emph{unfolding} a net, one has to unfold completely both \emph{time} and \emph{conflicts}, and this
yields a data structure that is generally infinite in time, and infinite also in conflicts (branchings) 
when choices are repeated. 

These difficulties have been addressed by limiting first the time dimension: 
the unfolding is restricted to a finite prefix 
(\cite{KKV:CPPU03}) that is sufficient (or ``complete'') to check the properties at stake, 
for example the reachability of some marking.
Still the resulting data structure may be unnecessarily big, and in the last decade, merged processes 
(\cite{KKKV:Acta06}) and the closely related trellis processes (\cite{Fab:Trellis}) were introduced to limit 
the expansion of the structure due to conflicts. The idea consists in merging runs that result from different 
choices but produce \emph{identical} resources, where identical may mean that the
same resource is produced by several alternative activities at the same time (trellis processes) or the
$i$-th occurrence of the same resource is produced by again alternative activities (merged processes).
These two approaches combined are quite successful to represent in a compact manner a sufficient set of runs 
of a concurrent system. However, they rely on distinct treatments for time and for conflicts. 

Spread nets are nets where each place is annotated, and the annotation depends on the transitions
putting a token in that place. In this way it is possible to keep track of the way that place is
\emph{reached}. 

Based on this notion, in the present paper we propose the notion of spreading of nets as a unified approach to 
Petri net unfolding. While trellises and merged processes had abandoned the requirement that nodes should not be in self-conflict in the unfolding, the main move here is to abandon also the requirement that the unfolding should be a directed acyclic graph. In other words, we consider structures that partially unfold time, and then loop back to previously met resources. This parametric approach is flexible enough to partially or totally expand both conflicts and time, thus capturing previous constructions in a unified setting. It also assigns an equal treatment to time and conflicts. 
We consider structures that are just ordinary nets where places are annotated with
\emph{vector-clocks}, and these annotations gather all the information about time and conflicts.

The capability of folding time resembles the notion of concatenation introduced on non-sequential behaviors of nets, whereas the capability of folding conflicts can be considered similar to the so called \emph{collective tokens} interpretations in 
Petri nets. According to this interpretation, the way a token is produced does not influence the subsequent use of 
it. With the capability of folding both time and conflict we allow to have that certain components of the net are 
executed according the individual token philosophy, whereas other parts may have different interpretations.

\paragraph{Structure of the paper:}
The paper is organized as follows. In the next section we recall the basic definitions about nets and we introduce 
\emph{multi-clock} nets. In Section~\ref{sec:ticking-domains} we define the domains of information on which our spreading
strategy is based. Spread nets over a suitable domain of information 
are then presented in Section~\ref{sec:spread-nets}.
In Section \ref{sec:spreading alg} we first introduce the spreading operation of an multi-clock net, 
and then we show that the spreading of a net enjoys some nice algebraic properties similar to the ones of 
unfoldings and trellis processes. We also show that indeed
our spreading strategy covers the ones of trellis and branching processes.  

\section{Nets}\label{sec:mcn}
 \paragraph{Notation:}
 With $\nat$ we denote the set of natural numbers.  
 Let $X$ be a set, with $\card{X}$ we denote the cardinality of the set. 
 Let $A$ be a set, a {\em multiset\/} of $A$ is a function $f : A\to \nat$.
 The usual operations on multisets, like multiset union $+$ or multiset difference $-$, are 
 defined in the standard way.
 We write $f \leq f'$ if $f(a) \leq f'(a)$ for all $a \in A$.  
 If a multiset $f$ is a set, \emph{i.e.} for all $a \in A$. $f(a)\leq 1$, we confuse 
 the multiset with the set and write $a\in f$ to indicate that $f(a) = 1$.
 
 Given an alphabet $\Sigma$, with $\Sigma^{\ast}$ we denote as usual
 the set of words on $\Sigma$, and with $\varepsilon$ the empty word.  The
 length of a word is defined as usual and, with abuse of notation, it
 is denoted with $\card{\cdot}{}$.
 
 Given two mapping $f \colon A \to B$ and $g \colon B \to C$, with $f\circ g$ we denote
 the composition of the two mappings defined as $f \circ g(a) = g(f(a))$.
 
 \paragraph{(Safe) Nets:}
 We first review the notions of (safe) labeled Petri net and of the token game. Consider a set $\Sigma$ 
 of names. 
 \begin{definition}
   \label{de:petri net}
   A labeled \emph{Petri net} over $\Sigma$ is a 5-tuple $N = \langle
   \place, T, F, m, \ell, \Sigma\rangle$, where
   \begin{itemize}
     \item  $\place$ is a set of {\em places} and $T$ is a set of {\em transitions} 
          (with $\place \cap T = \emptyset$), 
     \item $F \subseteq (\place\times T)\cup (T\times \place)$ is the \emph{flow} relation, 
     \item $m \colon \place\to \nat$ is called the {\em initial marking}, and 
     \item $\ell \colon T\cup\place \to \Sigma$ is a labeling mapping.
   \end{itemize}
 \end{definition} 
 With respect to the usual definition we have already added the labeling mapping, which is defined
 both on places and transitions.
 This will be handy when defining spread nets.
 Clearly ordinary Petri nets are those where the labeling is the identity (thus the transition names 
 are the transitions themselves, and place names are the places themselves).
 Subscripts or superscripts on the net name carry over to the names of the net components.
 Given $x\in T\cup\place$, $\pre{x} = \setcomp{y}{(y,x)\in F}$ and
 and $\post{x} = \setcomp{y}{(x,y)\in F}$. 
 $\pre{x}$ and $\post{x}$ are called the \emph{preset} and \emph{postset} respectively of $x$.
 Observe that, given a $t\in T$, $\pre{t}$ and $\post{t}$ can be seen ad multisets over $\place$, as well
 as a marking $m$.
 A net $\langle \place, T, F, m, \ell, \Sigma\rangle$ is as usual graphically represented  
 as a bipartite directed graph where the nodes are the places and the transitions, and where
 an arc connects a place $p$ to a transition $t$ iff $(p,t) \in F$ and 
 an arc connects a transition $t$ to a place $p$ iff $(t,p) \in F$. 
 We assume that all nets we consider are such that $\forall t\in T$ $\pre{t}$ and $\post{t}$ are not
 empty.
 
 A transition $t$ is enabled
 at a marking $m$, if $m$ \emph{contains} the preset of $t$, where contain here means that $m(p)\geq 1$ for
 all $p\in\pre{t}$, or equivalently $\pre{t}\leq m$. 
 If a transition $t$ is enabled at a marking $m$ it may \emph{fire} yelding a new marking defined as 
 $m'(p) = m(p) - \card{\pre{t}\cap\setenum{p}} + \card{\post{t}\cap\setenum{p}}$, or equivalently
 $m' = m -\pre{t} + \post{t}$. The firing of $t$ at 
 $m$ giving $m'$ is denoted as $m\trans{t} m'$. 
 The set of reachable markings of a net $N$ is denoted with
 $\mathcal{M}_{N}$. 
 A net $N$ is said to be \emph{safe} whenever its places hold at most one token in all possible 
 reachable marking, namely  $\forall m\in\mathcal{M}_{N}$ it holds that $m$ can be seen as a set (the only
 possible values are $0$ and $1$). 
 As markings may be considered as the characteristic function of a set, we will often confuse markings with 
 subsets of places.

\paragraph{Net morphisms:}
 We recall now the notion of morphism between safe nets~\cite{Win:ES}. 
 \begin{definition}
   \label{de:net morphism}
   Let $N = \langle \place, T, F, m, \ell, \Sigma\rangle$ and 
   $N' = \langle \place', T', F', m', \ell', \Sigma'\rangle$ be safe nets over $\Sigma$ and $\Sigma'$ respectively. 
   A {\em morphism} $\phi: N\to N'$ is the triple $\langle \phi_T, \phi_{\place}, \phi_{\ell} \rangle$, where
   \begin{itemize}
     \item   
        $\phi_T : T \rightarrow T'$ is a partial function and $\phi_{\place} \subseteq \place \times \place'$ is a
        relation such that
      \begin{itemize}
       \item 
          for each $p'\in m'$ there exists a unique $p\in m$ and $p\ \phi_{\place}\ p'$,     
       \item
          if $p\ \phi_{\place}\ p'$ then the restriction $\phi_T\colon \pre{p} \to \pre{p'}$ and 
          $\phi_T\colon \post{p} \to \post{p'}$ are total functions, and
       \item 
          if $t' = \phi_T(t)$ then $\phi_{\place}^{op}\colon \pre{t'}\to\pre{t}$ and 
          $\phi_{\place}^{op}\colon \post{t'}\to\post{t}$ are total
          functions, 
          where $\phi_{\place}^{op}$ is the opposite relation to $\phi_{\place}$, and 
       \end{itemize}
      \item 
         $\phi_{\ell} : \Sigma \to \Sigma'$ is such that if $\phi_T(t)$ is defined
         then $\ell'(\phi_T(t)) = \phi_{\ell}(\ell(t))$ and if $p\ \phi_{\place}\ p'$ then
        $\ell'(p') = \phi_{\ell}(\ell(p))$.        
   \end{itemize}
 \end{definition}
 The definition is the usual one, beside the last requirement which states that the labeling of the nets
 is preserved. We will omit the subscript when it will clear from the context, hence the triple
 $\langle \phi_T, \phi_{\place}, \phi_{\ell} \rangle$ will be often indicated as $\phi$.
 
 Morphisms among safe nets preserve reachable markings. Consider the morphism
 $\phi\colon N \to N'$, then for each
 $m, m' \in \mathcal{M}_{N}$ and transition $t\in T$, if
 $m \trans{t} m'$ then $\phi_{\place}(m) \trans{\phi_T(t)} \phi_{\place}(m')$ provided that $\phi_T(t)$
 is defined, where 
 $\phi_{\place}(m) = \{p'\in {\place}'\ |\ \exists p\in m\ \mathit{and}\ p\ \phi_{\place}\ p'\}$.
 
 Clearly morphisms compose and then safe nets and morphisms form a category called $\SafeNet$.

 \paragraph{Multi-clock nets:} 
 Safe nets can be seen as formed by various \emph{sequential} components (automata) 
 synchronizing on common transitions. 
 Though this is not the usual way to consider safe nets, it is easy to see that if we add to a 
 safe nets the so called \emph{complemenary} places, except in the case of self-loops, we obtain
 a number of automata synchronizing on common transitions.
 A net automaton is a net in which the 
 preset and the postset of each transition has exactly one element. 
 
 The intuition  that a safe net can be viewed as a net formed by various components, each of
 them being a net automaton, is formalized in the notion of 
 \emph{multi-clock} nets, introduced by Fabre in \cite{Fab:Trellis}.

 \begin{definition}
  \label{de:mc-net}
  A \emph{multi-clock} net (\mcn) $\mathsf{N}$ is a pair $(N, \partition)$ 
  where $N = \langle \place, T, F, m, \ell, \Sigma\rangle$ is a safe net and $\partition : \place \to m$ is
  a mapping such that
  \begin{itemize}
    \item
      for all $p, p'\in m$, it holds that $p\neq p'$ implies 
      $\partition^{-1}(p) \cap \partition^{-1}(p') = \emptyset$, 
    \item
      $\bigcup_{p\in m}\partition^{-1}(p) = \place$,
    \item 
      $\partition$ is the identity when restricted to $m$, and
    \item 
      for all $t\in T$. $\partition$ is injective on $\pre{t}$ and on 
      $\post{t}$, and $\nu(\pre{t}) = \nu(\post{t})$.  
  \end{itemize}
  The \emph{dimension} of a \mcn\ $\mathsf{N}$, denoted with $\ind{\mathsf{N}}$, is the
  cardinality of $m$.
 \end{definition}
 The mapping $\partition$ is used to identify the various components of a \mcn.
 Given $p\in P$, with $\overline{p}$ we denote the subset of places defined by $\partition^{-1}(\partition(p))$.  
 The consequences of three requirements, namely (a) $\partition(m) = m$, (b) 
 $\partition$ is injective on the preset (postset) of each transition and (c) that 
 $\partition(\pre{t}) = \partition(\post{t})$, is that, for each
 $p\in m$, the net
 $\langle \overline{p}, T_{\overline{p}}, F_{\overline{p}}, \setenum{p}, \ell_{\overline{p}}, \Sigma\rangle$ 
 is a net automaton, where 
 $T_{\overline{p}}$ are the transitions of $N$ such that $\forall t\in T_{\overline{p}}$ 
 $\pre{t}\cap \overline{p} \neq \emptyset$ and $\post{t}\cap \overline{p} \neq \emptyset$, and $F_{\overline{p}}$
 is the restriction of $F$ to ${\overline{p}}$ and $T_{\overline{p}}$.
 Each place $p$ in the initial marking can be identified with an 
 index in $\setenum{1, \dots, \ind{\mathsf{N}}}$, hence
 we denote $N_i$ as the net  
 $\langle \overline{p}, T_{\overline{p}}, F_{\overline{p}}, \setenum{p}\ell_{\overline{p}}\rangle$ where
 $i$ is the \emph{index} of $p$. 
 Thus the cardinality of the initial marking of a \mcn\ is the number of components forming
 the net, and it is the dimension of the net.

 \begin{example}\label{ex:running-example}
 Consider the \mcn\ $\mathsf{N}$ in Figure~\ref{fig:mcn-running-example}. 
 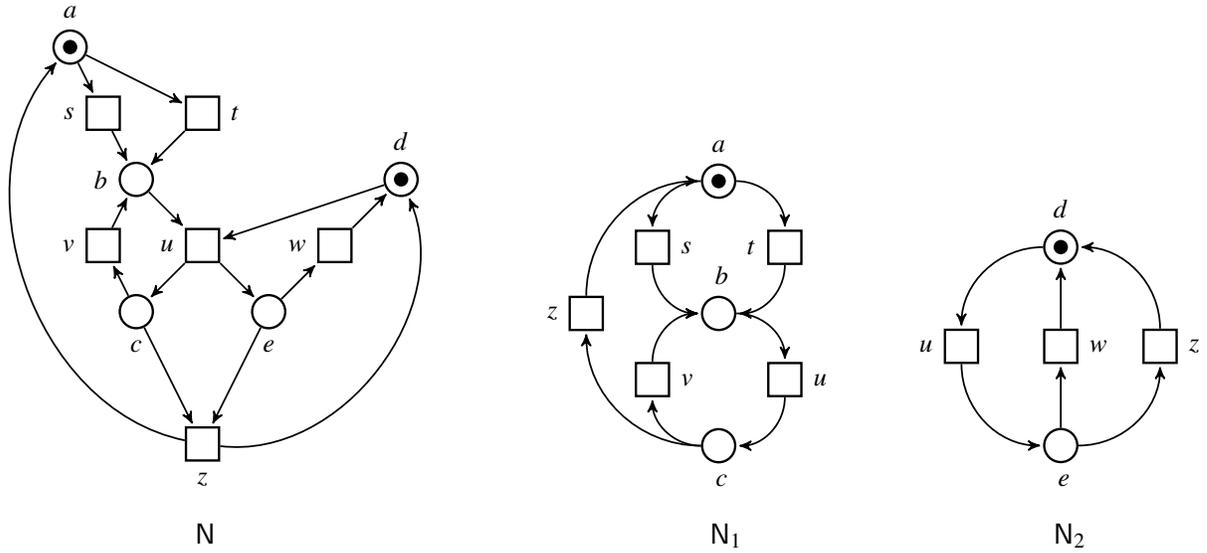
\begin{figure}[!ht]
 \centerline{
 \begin{tabular}{ccccc}
  & & & & \\
 \scalebox{1.1}{
\begin{tikzpicture}
[bend angle=60, scale=0.8, pre/.style={<-,shorten
    <=1pt,>=stealth',semithick}, post/.style={->,shorten
    >=1pt,>=stealth',semithick},place/.style={circle, draw=black,
    thick, minimum size = 4mm}, transition/.style={rectangle, draw=black, thick, minimum size = 4mm},
    posta/.style={-,semithick},
    invplace/.style={circle, draw=black!0,thick}]    
\node[place, tokens = 1] (a) at (0,5) [label = above:\footnotesize{$a$}]{};
\node[place] (c) at (1,1) [label = below:\footnotesize{$c$}]{};
\node[place] (b) at (1,3)[label = left:\footnotesize{$b$}]{};
\node[place, tokens = 1] (d) at (5,3) [label = above:\footnotesize{$d$}]{};
\node[place] (e) at (3,1) [label = below:\footnotesize{$e$}]{};
\node[transition] (s) at ( .5, 4 ) [label =left:\footnotesize{$s$}]{}
edge[pre] (a)
edge[post] (b);
\node[transition] (t) at ( 2, 4 ) [label =right:\footnotesize{$t$}]{}
edge[pre] (a)
edge[post] (b);
\node[transition] (u) at ( 2, 2 ) [label = left:\footnotesize{$u$}]{}
edge[pre] (b)
edge[pre] (d)
edge[post] (c)
edge[post] (e);
\node[transition] (v) at ( .5, 2 ) [label = left:\footnotesize{$v$}]{}
edge[pre] (c)
edge[post] (b);
\node[transition] (w) at ( 4, 2 ) [label = left:\footnotesize{$w$}]{}
edge[pre] (e)
edge[post] (d);
\node[transition] (z) at ( 2, -1) [label = below:\footnotesize{$z$}]{}
edge[pre] (c)
edge[pre] (e)
edge[post, bend left] (a)
edge[post, bend right] (d);
\end{tikzpicture}
}
&
\quad
&
\scalebox{1.1}{
\begin{tikzpicture}
  [bend angle=45, scale=0.8, pre/.style={<-,shorten
    <=1pt,>=stealth',semithick}, post/.style={->,shorten
    >=1pt,>=stealth',semithick},place/.style={circle, draw=black,
    thick, minimum size = 4mm}, transition/.style={rectangle, draw=black, thick, minimum size = 4mm},
    invplace/.style={circle, draw=black!0,thick}]
\node[invplace] (inv) at (1,0.7){};
\node[place] (b) at (1,3)[label = above:\footnotesize{ $b$}]{};
\node[place] (c) at (1,1)[label = below:\footnotesize{ $c$}]{};
\node[place, tokens = 1] (a) at (1,5) [label = above:\footnotesize{$a$}]{};
\node[transition] (s) at ( 0, 4 ) [label =right:\footnotesize{$s$}]{}
edge[pre, bend left] (a)
edge[post, bend right] (b);
\node[transition] (t) at ( 2, 4 ) [label =left:\footnotesize{$t$}]{}
edge[pre, bend right] (a)
edge[post, bend left] (b);
\node[transition] (u) at ( 2, 2 ) [label = right:\footnotesize{$u$}]{}
edge[pre, bend right] (b)
edge[post, bend left] (c);
\node[transition] (v) at ( 0, 2 ) [label = right:\footnotesize{$v$}]{}
edge[pre, bend right] (c)
edge[post, bend left] (b);
\node[transition] (z) at ( -1, 3) [label = left:\footnotesize{$z$}]{}
edge[pre, bend right] (c)
edge[post, bend left] (a);
\end{tikzpicture}}
&
\quad
&
\scalebox{1.1}{
 \begin{tikzpicture}
  [bend angle=45, scale=0.8, pre/.style={<-,shorten
    <=1pt,>=stealth',semithick}, post/.style={->,shorten
    >=1pt,>=stealth',semithick},place/.style={circle, draw=black,
    thick, minimum size = 4mm}, transition/.style={rectangle, draw=black, thick, minimum size = 4mm},
    invplace/.style={circle, draw=black!0,thick}]
\node[invplace] (inv) at (2,0.7){};
\node[place] (e) at (1,.5)[label = below:\footnotesize{ $e$}]{};
\node[place, tokens = 1] (d) at (1,3.5)[label = \footnotesize{$d$}]{};
\node[transition] (u) at ( -.5, 2 ) [label = left:\footnotesize{$u$}]{}
edge[pre, bend left] (d)
edge[post, bend right] (e);
\node[transition] (w) at ( 1, 2 ) [label = right:\footnotesize{$w$}]{}
edge[pre] (e)
edge[post] (d);
\node[transition] (z) at ( 2.5, 2) [label = right:\footnotesize{$z$}]{}
edge[pre, bend left] (e)
edge[post, bend right] (d);
\end{tikzpicture}} \\ [4pt]
 $\mathsf{N}$ & & \quad\qquad$\mathsf{N}_1$ & & \quad$\mathsf{N}_2$ \\
 \end{tabular}}
\caption{A \mcn\ and its components}\label{fig:mcn-running-example}
\end{figure}
 
 The $\partition$ in the \mcn\
 $\mathsf{N}$ gives
 $\partition(a) = \partition(b) = \partition(c) = \setenum{a}$ and $\partition(d) = \partition(e) = \setenum{d}$. 
 The two net automata are $\mathsf{N}_1$ and $\mathsf{N}_2$.
 The composition of the two  automata
 (identifying the transitions with the same name, namely $u$ and $z$) gives precisely
 $\mathsf{N}$.
 \end{example}
 
 We consider morphisms that preserve the partitions of multi-clock nets. 

 \begin{definition}
   \label{de:mcn morphism}
   Let $N = (\langle \place, T, F, m, \ell, \Sigma\rangle,\partition)$ and 
   $N' = (\langle \place', T', F', m', \ell', \Sigma'\rangle,\partition')$ be two multi-clock nets. 
   A {\em morphism} $\phi : N\to N'$ is a mcn-morphism iff 
   $\forall p\in \place$, $\forall p'\in \place'$, $p\ \phi_{\place}\ p'$ 
   implies that $\partition(p) \ \phi_{\place}\ \partition'(p')$.
 \end{definition}
 Multi-clock nets and mcn-morphisms form a category called $\MultiClockNet$, which is a subcategory of
 $\SafeNet$. 

 \begin{example}
  Consider the \mcn{s} $\mathsf{N}$ and $\mathsf{N}_1$ in Figure~\ref{fig:mcn-running-example}. A mcn-morphism is
  the one relating places in  $\mathsf{N}$ to places with the same name in  $\mathsf{N}_1$. 
  Places $d$ and $e$ in $\mathsf{N}$ are not related with any place in $\mathsf{N}_1$. 
  The mapping on the transitions is the identity on $s, t, u, v$ and $z$ and it is undefined for $w$.
 \end{example}
 
 In this paper, for the sake of simplicity, we will \emph{spread} \mcn{s} that are 
 \emph{injectively} labeled.

\section{Ticking domains}\label{sec:ticking-domains}
We introduce the \emph{annotations} for places of the spread nets, which we will define in the next section. 
Annotations will be vector-clocks, where each entry of the vector will be an equivalence class of words representing
the local runs of each component. This simple annotation will turn out to be powerful enough to represent most of
the situations we are interested in.

Consider an alphabet $\mathit{A}$, the idea is that the elements of a ticking 
domain are equivalence classes of words on that alphabet.
Let $\mathit{Eq}$ be a set of equalities of the form $u_i = u'_i$, with $u_i, u'_i\in \mathit{A}^{\ast}$. 
We denote by $\sim_{\mathit{Eq}}$ the equivalence relation in $\mathit{A}^{\ast}$ generated by 
relations in $\mathit{Eq}$ and that is stable by suffix extension, 
i.e. $u\sim_{\mathit{Eq}} u' \;\Rightarrow\; uv\sim_{\mathit{Eq}} u' v$ for $u,u',v\in\mathit{A}^{\ast}$. 
Relation  $\sim_{\mathit{Eq}}$ will simply be denoted $\sim$ when the generating set is 
clear from the context, and equivalence classes are denoted as $\eqclass{w}{\sim}$ or simply $\eqclass{w}{}$. 
For the sake of light notations, in the sequel we will often confuse $w$ with its class $\eqclass{w}{}$. 
Observe that one has  $\eqclass{\,\eqclass{u}{}v\,}{} =\eqclass{uv}{}$.

 \begin{definition}
  \label{de:tick-dom}
  Let $\mathit{A}$ be an alphabet and let $\mathit{Eq}$ be a set of equalities of the form
  $\alpha = \beta$, with $\alpha, \beta\in \mathit{A}^{\ast}$. Then a \emph{ticking domain} over
  $\mathit{A}$ is the set of equivalence classes of words in $\mathit{A}^{*}$ with respect to the suffix 
  stable equivalence relation
  $\sim_{\mathit{Eq}}$ generated by $\mathit{Eq}$, and it is denoted 
  $\mathcal{A}_{\mathit{Eq}}=\mathit{A}^{\ast}/\sim_{\mathit{Eq}}$.
  With $\mathsf{alph}(\mathcal{A}_{\mathit{Eq}})$ we denote the alphabet $\mathit{A}$. 
 \end{definition}
 
\begin{example}\label{ex:tick-running-example}
 Consider the \mcn{}  $\mathsf{N}_1$ in Figure~\ref{fig:mcn-running-example}. 
 The alphabet can be considered the name of the 
 transitions (hence $\mathit{A} = \setenum{s,t,u,v,z}$)
 and we may imagine the following equations: $s = t$, $suvs = s$ and $suzs = s$. These equations induce, 
 among others, the following equivalence classes on 
 the word representing some executions
 of the \mcn{}  $\mathsf{N}_1$: 
 $\eqclass{\varepsilon}{\sim_1}$, $\eqclass{s}{\sim_1}$, $\eqclass{su}{\sim_1}$, 
 $\eqclass{suv}{\sim_1}$ and $\eqclass{suz}{\sim_1}$ (the other equivalence 
 classes may be ignored, as it will become clear in
 the following). These equivalence classes form a ticking domain for the state-machine net $\mathsf{N}_1$.
 
 Another set of equations over $\mathit{A} = \setenum{s,t,u,v,z}^{*}$ can be the following: for all 
 $w, w' \in \setenum{s,t,u,v,z}$, $w = w'$ iff $\card{w} = \card{w'}$. In this case
 two words are in the same equivalence class iff they have the same length. 
 Requiring that the equivalence of words is also a congruence, the same set of equivalence classes could be obtained 
 from the set of equations $u = v$ with $u, v \in \setenum{s,t,u,v,z}$.
 
 If the set of equations is empty then 
 each word $w\in\mathit{A}^{\ast}$ is the unique member of the equivalence class $\eqclass{w}{}$. Being the 
 equivalence relation stable with respect to suffix, the same can be obtained using as the set of equations
 $u = u$ with $u \in \setenum{s,t,u,v,z}$.
\end{example}

 Given two ticking domains $\mathcal{A}_{\mathit{Eq}}$ and $\mathcal{A}_{\mathit{Eq}'}'$, 
 $\delta \colon \mathcal{A}_{\mathit{Eq}}\to \mathcal{A}'_{\mathit{Eq}'}$ is a ticking domain mapping
 iff given any two words $w, w'\in \mathit{A}^{*}$ in the same equivalence class in $\mathcal{A}_{\mathit{Eq}}$, then
 $\delta(w), \delta(w')$ are in the same equivalence class in $\mathcal{A}'_{\mathit{Eq}'}$.

 We are now ready to introduce the notion of \emph{vector-clock}.
 \begin{definition}
  \label{de:vector-clock}
  Given a set of index $I$ and a set of ticking domains $\mathcal{A}_i$, with $i\in I$, 
  a \emph{vector-clock} $\overrightarrow{\alpha}$ is an element
  of $\times_{i\in I} \mathcal{A}_i$, and $\mathcal{A} = \times_{i\in I} \mathcal{A}_i$ is 
  called the \emph{vector-clock domain} (\vcd\ for short). The \emph{dimension} of the vector-clock domain
  $\mathcal{A}$, denoted with $\dime{\mathcal{A}}$, is given by $\card{I}$.
 \end{definition}
 The $\times$ on clock domains is associative and can be easily extended to an operation $\times$ on vector-clock
 domains as a component-wise operation.
 
 Vector clock elements can be \emph{mixed} to obtain a new vector clock element. The intuition is that the
 new element is obtained from the previous one selecting entries from each of them. This is formally stated in the
 next definition.
 
 \begin{definition}
  \label{de:op}
   Let $J$ be a set of index and let $\Gamma = \setcomp{\alpha_j}{j\in J\ \land\ \alpha_j\in\mathcal{A}}$ 
   be a set of vector clock
   in $\mathcal{A}$. 
   Then
   $\mathit{op}_{J}^{k} \colon \mathcal{A}^{\card{J}} \to \mathcal{A}$, with $k\in J$, is an operation defined as
   follows: the $i$-th entry of $\mathit{op}_{J}^{k}(\Gamma)$ is the $i$-th entry of the vector 
   clock $\alpha_i\in \Gamma$ if
   $i\in J$ and of $\alpha_k\in\Gamma$ otherwise.
 \end{definition}
 We briefly discuss the intuition behind this definition. The various components of a vector-clock represent
 the pieces of information each component has on its behavior and on the other components behaviors as well.
 The various pieces of information have to be combined together to form a new vector clock, which will be the
 argument of a function of a spread net. 
 The way of combining the information should take into account mainly the information associated to a certain
 set of indexes, as it will be again clear when we will introduce the notion of spread net.
 
 Clearly these pieces of information have to be consistent, and the operations defined above 
 are responsible in assuring this. 
 
 \begin{example}
  Take $\Gamma = \setenum{(w^{j_1}_1, w^{j_1}_2, w^{j_1}_3), (w^{j_2}_1, w^{j_2}_2, w^{j_2}_2)}$, and 
  $\mathit{op}_{J}^{j_1}$, where $J = \setenum{j_1, j_2}$. Then $\mathit{op}_{J}^{j_1}(\Gamma)$ is
  $(w^{j_1}_1, w^{j_2}_2, w^{j_1}_3)$, whereas $\mathit{op}_{J}^{j_2}(\Gamma)$ is
  $(w^{j_1}_1, w^{j_2}_2, w^{j_2}_3)$.
 \end{example}

\section{Spread Nets}\label{sec:spread-nets}
 We enrich \mcn{s} with vector clocks. 
 The idea is that
 each place of a \mcn{} $\mathsf{N}$ of dimension $\ind{\mathsf{N}}$
 has associated a vector-clock belonging to a vector clock domain $\mathcal{A}$ such that 
 $\dime{\mathcal{A}} = \ind{\mathsf{N}}$.
 Thus in general the annotation of a place of a \mcn{} carries information on the component the place 
 belongs to (the proper entry in the vector clock), but it
 may also convey information about the other components (the other entries of the vector clock).
 
 We start by illustrating this idea with a little example.
 
 \begin{example}\label{ex:spread-int-1}
 Consider the \vcd\ 
 $\mathcal{A}_1 \times \mathcal{A}_2 \times\mathcal{A}_3$, where  
 $\mathcal{A}_1 = \{\eqclass{\varepsilon}{\sim_1},$ $\eqclass{v}{\sim_1},\eqclass{u}{\sim_1},\eqclass{us}{\sim_1}\}$,
 with $\sim_1$ obtained by the equations $uu = u$, $usv = us$,  
 $usu = us$, $uss = us$, $vs = v$, $vu = v$, $vv = v$ and  $s = \varepsilon$,  
 $\mathcal{A}_2 = \setenum{\eqclass{\varepsilon}{\sim_2},\eqclass{u}{\sim_2},\eqclass{us}{\sim_2}}$,
 with $\sim_2$  induced by the set of equations $\mathit{Eq}_2 = \setenum{u = w,$ $us = ws,$ 
 $s = \varepsilon,$ $su = \varepsilon, sw = \varepsilon, uu = u, uw = u}$, and 
 $\mathcal{A}_3 = \setenum{\eqclass{\varepsilon}{\sim_3}}$, with 
 $\sim_3$ induced from the set $\mathit{Eq}_3 = \setenum{\varepsilon = w}$. 
  
 Consider now the \mcn{} in Figure~\ref{fig:spread-example}.

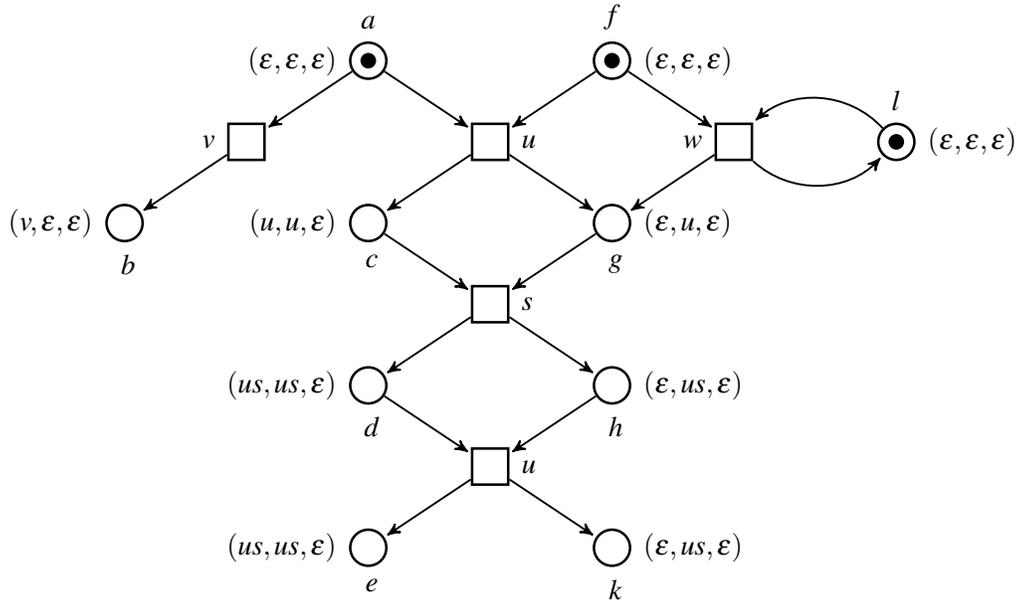
\begin{figure}[!t]
\centerline{
\scalebox{1.2}{\begin{tikzpicture}
[bend angle=45, scale=0.9, pre/.style={<-,shorten
    <=1pt,>=stealth',semithick}, post/.style={->,shorten
    >=1pt,>=stealth',semithick},place/.style={circle, draw=black,
    thick, minimum size = 4mm}, transition/.style={rectangle, draw=black, thick, minimum size = 4mm},
    invplace/.style={circle, draw=black!0,thick}]
\node[invplace] (inv) at (1,4){};
\node[invplace] (inv) at (1,-1.5){};
\node[place, tokens = 1] (s1) at (1,3)[label = above:\footnotesize{$a$},
                                       label = left:\footnotesize{$(\varepsilon,\varepsilon,\varepsilon)$}]{};
\node[place, tokens = 1] (s2) at (4,3)[label = above:\footnotesize{$f$},
                                       label = right:\footnotesize{$(\varepsilon,\varepsilon,\varepsilon)$}]{};
\node[place, tokens = 1] (s3) at (7.5,2)[label = above:\footnotesize{$l$},
                                       label = right:\footnotesize{$(\varepsilon,\varepsilon,\varepsilon)$}]{};
\node[place] (r1) at (-2,1)[label = below:\footnotesize{ $b$},
                                       label = left:\footnotesize{$(v,\varepsilon,\varepsilon)$}]{};
\node[place] (r2) at (1,1)[label = below:\footnotesize{ $c$},
                                       label = left:\footnotesize{$(u,u,\varepsilon)$}]{};
\node[place] (r3) at (4,1)[label = below:\footnotesize{ $g$},
                                       label = right:\footnotesize{$(\varepsilon,u,\varepsilon)$}]{};
\node[place] (p1) at (1,-1)[label = below:\footnotesize{ $d$},
                                       label = left:\footnotesize{$(us,us,\varepsilon)$}]{};
\node[place] (p2) at (4,-1)[label = below:\footnotesize{ $h$},
                                       label = right:\footnotesize{$(\epsilon,us,\varepsilon)$}]{};
\node[place] (q1) at (1,-3)[label = below:\footnotesize{ $e$},
                                       label = left:\footnotesize{$(us,us,\varepsilon)$}]{};
\node[place] (q2) at (4,-3)[label = below:\footnotesize{ $k$},
                                       label = right:\footnotesize{$(\epsilon,us,\varepsilon)$}]{};
\node[transition] (b) at ( 2.5, 2 ) [label = right:\footnotesize{$u$}]{}
edge[pre] (s2)
edge[pre] (s1)
edge[post] (r2)
edge[post] (r3);
\node[transition] (s) at ( 2.5, 0 ) [label = right:\footnotesize{$s$}]{}
edge[pre] (r2)
edge[pre] (r3)
edge[post] (p1)
edge[post] (p2);
\node[transition] (c) at ( -.5, 2 ) [label = left:\footnotesize{$v$}]{}
edge[pre] (s1)
edge[post] (r1);
\node[transition] (a) at ( 5.5, 2 ) [label = left:\footnotesize{$w$}]{}
edge[pre] (s2)
edge[pre, bend left] (s3)
edge[post, bend right] (s3)
edge[post] (r3);
\node[transition] (b1) at ( 2.5, -2 ) [label = right:\footnotesize{$u$}]{}
edge[pre] (p2)
edge[pre] (p1)
edge[post] (q2)
edge[post] (q1);
\end{tikzpicture}}}
\caption{A net over an information domain}\label{fig:spread-example}
\end{figure}
The components of this \mcn{} are identified by the partition mapping defined as follows: 
$\partition^{-1}(a) = \setenum{a, b, c, d, e}$, $\partition^{-1}(f) = \setenum{f, g, h, k}$ and
$\partition^{-1}(l) = \setenum{l}$. For 
the ticking domains $\mathcal{A}_1, \mathcal{A}_2$ and $\mathcal{A}_3$ the alphabets on which they are based 
is the set of transitions of each component, thus $\mathsf{alph}(\mathcal{A}_1) = \setenum{v, u, s}$, 
$\mathsf{alph}(\mathcal{A}_2) = \setenum{u, s, w}$ and $\mathsf{alph}(\mathcal{A}_2) = \setenum{w}$.
The vector clocks associated to the places of this \mcn{} are the
ones shown in the figure: to $a, f$ and $l$ the vector clock associated is $(\varepsilon, \varepsilon, \varepsilon)$,
to place $b$ the vector clock $(v,\varepsilon,\varepsilon)$, to $c$ the vector clock $(u,u,\varepsilon)$, to
$f$ the vector-clock $(u,u,\varepsilon)$, to $d$ and $e$ the $(us,us,\varepsilon)$ and finally
to $h$ and $k$ the vector-clock $(\varepsilon,us,\varepsilon)$.
\end{example}

\paragraph{\mcn{} over $\mathcal{A}$:} 
We first introduce the notion of \mcn{} over a domain and then we will formalize the one of spread net. 

\begin{definition}
  \label{de:nets over domains}
  The pair 
  $\mathcal{N} = ((\langle\place, T, F, m, \ell, \Sigma\rangle, \partition),  
   h \colon \place\to \mathcal{A})$, 
  where 
  \begin{itemize}
    \item $\mathsf{N} = (\langle\place, T, F, m, \ell, \Sigma\rangle, \partition)$ is a \mcn, 
    \item $\mathcal{A}$ is a vector clock domain such that 
          $\mathcal{A} = \times_{i=1}^{\ind{\mathsf{N}}}{\mathcal{A}_i}$, where 
          $\mathsf{alph}(\mathcal{A}_i) = \ell(T_i)$
          with $1\leq i\leq \ind{\mathsf{N}}$, and
    \item $h \colon \place \to  \mathcal{A}$ is a total mapping and it is called the \emph{information} mapping.
  \end{itemize}
  is called a \mcn{} over $\mathcal{A}$.
  With $\mathsf{td}(\mathcal{N})$ we denote the vector clock domain $\mathcal{A}$.
 \end{definition}
 
 \paragraph{Spread nets:}
 We assume, for each $\mathcal{A}_i$, that the set of equations $\mathit{Eq}_i$ 
 on words over the alphabet $\mathsf{alph}(\mathcal{A}_i)^{\ast}$ such that $\mathcal{A}_i =
 \mathsf{alph}(\mathcal{A}_i)^{\ast}/\sim_{\mathit{Eq}_i}$, is well understood. 
 A \mcn{} over a \vcd{} is then an annotated net, where the annotations are on places. 
 These annotations are of a specific kind, namely they are vector-clocks
 where each component of the vector is an equivalence class of words on given alphabets.
 Based on this notion we can introduce the notion of \emph{spread} net, where the annotations of 
 places are \emph{calculated}.
 
 \begin{definition}
  \label{de:spread nets}
  Let $\mathsf{N}$ be a \mcn{} and $\mathcal{A}$ be a vector clock domain such that
  $\dime{\mathcal{A}} = \ind{\mathsf{N}}$. 
  Let $\mathcal{S}=(\mathsf{N}, 
  h\colon\place \to\mathcal{A})$ be a \mcn{} over $\mathcal{A}$. 
  Let 
  $\vec{\tau} = \setcomp{\tau_i}{1\leq i\leq\ind{\mathsf{N}}}$ be a set of \emph{ticking} mapping with 
  $\tau_i : \mathcal{A}\times T_i \to \mathcal{A}$
  Then $\mathcal{S}$ is a \emph{spread} net with respect to $\vec{\tau}$ and $\mathcal{A}$
  iff
  \begin{itemize}
  \item $\forall p.\ m(p) = 1$ it holds that $h(p) = (\varepsilon, \dots, \varepsilon)$,
  \item $\forall p, p'\in \place$.\ $\ell(p) = \ell(p')$ and $h(p) = h(p')$ implies $p = p'$, and 
  \item $\forall t\in T$. $\forall p\in\post{t}$ 
        $h(p) = \tau_i(\mathit{op}_{\partition(\pre{t})}^{\partition(p)}(\setcomp{h(p')}{p'\in\pre{t}}),t)$.
  \end{itemize} 
  With $\mathsf{support}(\mathcal{S})$ we denote the \mcn{} $\mathsf{N}$.
 \end{definition}
A spread net is a \mcn{} over a specific domain where the annotations on the places in the
postset of a transition $t$ are related to annotations of the places in the preset of this transition and
the transition $t$ itself. 
The annotations are a mean to keep track on how a place can be \emph{reached}. Thus
we require that the annotations of the places in the initial marking is $(\varepsilon, \dots, \varepsilon)$, and
the annotations on the places in the postset of a transition are calculated on the basis of the annotations
in the preset of this transition (combined using the operations according to Definition~\ref{de:op}):
the annotations of the places in the component $i$ 
are \emph{calculated} using a function $\tau_i$ which is \emph{local} to the component itself, though the
$\mathit{op}_{\partition(\pre{t})}^{\partition(p)}$ may not be local at all.
The requirement that $\forall p, p'\in \place$.\ $\ell(p) = \ell(p')$ and $h(p) = h(p')$ implies $p = p'$
implies that if two equally labeled places have the same information, then they are indeed the same place.
Indeed two equally labeled places represent the same activity and if the annotation is the same then, despite
the various possible alternatives that may have produced them, they should not be distinguished and hence are the 
same. 
This is a \emph{succinctness} principle that avoid that the same information is associated to different places
representing the same resource. 

\begin{example}\label{ex:spread-int-2}
Consider again the net in Figure~\ref{fig:spread-example}. 
Assume that the alphabets for the various components are the transition labels (in this case, as
the net is injectively labeled, these coincide with the transitions themselves). 

Take $\tau_1$ as the mapping that concatenatee the transition
$t$ to each entry $i$ of the vector clock such that 
$t\in \mathsf{alph}(\mathcal{A}_i)$ and leave the other entries untouched,
$\tau_2$ adds the transition $t$ just to the second component and finally $\tau_3$ is the constant mapping
giving $(\varepsilon,\varepsilon,\varepsilon)$. 

When executing each transition of the net we assume that the vector clocks associated
to the place in the postset of a transition are \emph{calculated} from the vector clocks associated
to the places in the preset of this transition, and that this is done \emph{locally}. Thus, when executing
$v$, the vector clock $(\varepsilon,\varepsilon,\varepsilon)$ of $a$ is used to obtain $(v,\varepsilon,\varepsilon)$ on $b$,
or when executing $u$ the vectors clocks $(\varepsilon,\varepsilon,\varepsilon)$ associated to $a$ and
$(\varepsilon,\varepsilon,\varepsilon)$ associated to $e$ are used to obtain 
$(u,u,\varepsilon)$ for $c$ and $(\varepsilon, w, \varepsilon)$ for $f$, recalling that $u \sim_2 w$.
The two vector clocks in $\pre{u}$ are \emph{merged} together, by selecting the proper components.
The vector-clocks associated to $g$ and $d$ are obtained first calculating a vector clock from the one in $\pre{s}$,
and $(u,u,\varepsilon)$ and $(\varepsilon,u,\varepsilon)$ are merged obtaining $(u,u,\varepsilon)$ and then
$d$ gets $(us,us,\varepsilon)$ whereas $g$ gets the vector clock $(\varepsilon,us,\varepsilon)$.
\end{example}
 
\paragraph{Morphisms:} We specialize to this new setting the notion of morphism:
 \begin{definition}
   \label{de:net over domains morphism}
   Let $\mathcal{S} = (\mathsf{N}, h \colon\place\to\mathcal{A})$ and 
   $\mathcal{S}' = (\mathsf{N}',  h' \colon\place'\to\mathcal{A}')$ be two spread nets, 
   $\mathcal{S}$ over $\vec{\tau}$ and $\mathcal{A}$ and
   $\mathcal{S}'$ over $\vec{\tau}'$ and $\mathcal{A}'$ respectively. 
   A \emph{spread}-morphism $f: \mathcal{S} \to \mathcal{S}$ is a pair $f = (\phi, \delta)$ where 
   \begin{itemize}
     \item $\phi \colon \mathsf{N} \to \mathsf{N'}$ is
           a mcn-morphism,
    \item $\delta : \mathcal{A} \to \mathcal{A}'$ is a mapping such that for each $\tau_i\in\vec{\tau}$ 
          and $\tau'_i\in \vec{\tau}'$ it holds that 
          $\tau'_i(\delta(\alpha),\phi_T(t)) = \delta(\tau_i(\alpha,t))$ whenever $\phi_T(t)$ is defined
           and $t\in T_i$, and
    \item $\delta(h(p)) = h'(p')$ whenever $p\ \phi_{\place}\ p'$.       
   \end{itemize}    
 \end{definition}
 
 We show that spread-morphisms compose. Let $f = (\phi,\delta) : \mathcal{S} \to \mathcal{S}'$ and
 $g = (\phi',\delta') : \mathcal{S}' \to \mathcal{S}''$ two \netdom-morphisms. 
 $f\circ g : \mathcal{S}\to \mathcal{S}''$ defined as 
 $(\phi\circ\phi', \delta\circ\delta')$ is a well defined
 \netdom-morphism. 
 The only condition to check is that $\delta\circ\delta'(\tau_i(\alpha,t))$ is defined
 whenever also $\phi_T\circ\phi'_T(t)$ is defined, and it is equal to 
 $\tau_i''(\delta\circ\delta'(\alpha),\phi_T\circ\phi'_T(t)(t))$. 
 Now $\delta\circ\delta'(\tau_i(\alpha,t)) = \delta'(\delta(\tau_i(\alpha,t)))$ and this is equal to
 $\delta'(\tau_i'(\delta(\alpha),\phi_T(t)))$ and finally also to 
 $\tau_i''(\delta'(\delta(\alpha)),\phi'_T(\phi_T(t)))$ which is
 $\tau_i(\delta\circ\delta'(\alpha),\phi_T\circ\phi'_T(t))$ as required. 
 Clearly we have that $\delta\circ\delta'(h(p)) = \delta'(\delta(h(p))) = \delta'(h(p'))
 = h'(p'')$ with $p\ \phi_{\place}\ p'$ and $p'\ \phi_{\place}\ p''$.
 Thus spread nets and spread-morphisms form a category, that we call $\Spread$. 
 
 This category is related to the one of \mcn{s} via two obvious functors. One takes an
 object $\mathcal{S}$ in $\Spread$ and returns the \mcn{} $\mathsf{support}(\mathcal{S})$, and we
 call it $\mathfrak{F}$, the other takes a \mcn{} $\mathsf{N}$ where the labeling $\ell$ is injective 
 and associate the spread net
 $\mathfrak{G}(\mathsf{N})$ over $\vec{\tau} = \{\tau_i\ |\ \tau_i$ is the constant mapping returning   
 $(\varepsilon,\dots,\varepsilon)\}$ 
 and $\mathcal{A}_{\bot}$ defined as $(\mathsf{N}, h\colon \place\to \setenum{(\varepsilon,\dots,\varepsilon)})$
 (thus all the places are annotated with the vector-clock
 $(\varepsilon,\dots,\varepsilon)$), and each mcn-morphism $\phi$ gives 
 $\mathfrak{G}(\phi) = (\phi,\mathit{id})$.
 
 When it will be clear from the context, we will omit to mention both the $\vec{\tau}$ and $\mathcal{A}$ 
 on which a spread net is based on.

\paragraph{Configuration:}
 We end this section by defining what a \emph{configuration} of a spread net is. This notion will be used when
 spreading a net, and it is the usual one adapted to the context of spread nets.

 \begin{definition}\label{de:configuration}
 Let 
 $\mathcal{S} = (\mathsf{N}, h \colon\place\to\mathcal{A})$ be a 
 spread net, and
 $m\trans{t_1}m_1 \dots n_{n-1}\trans{t_n}m_n$ be a firing sequence in $\mathsf{N}$. Then a configuration is
 the multiset $\sum_{i=1}^{n} \setenum{t_i}$. 
 \end{definition}
  
 Configurations are ranged over with $C$ and
 $m_n = \mathit{mark}(C)$ is marking reached executing the firing sequence associated to the configuration.
 The set of configuration of a spread net $\mathcal{S}$ is denoted with $\conf{\mathcal{S}}{}$.

\section{Spreading nets}\label{sec:spreading alg}
In this section we describe how to spread \mcn{s}. We assume that the labeling mapping of the net that should be
spread is the identity.

First we recall what a \emph{folding}-morphism is.
Let $\mathsf{N}$ and $\mathsf{N}'$
be \mcn{s}. 
$\phi \colon\mathsf{N} \to \mathsf{N}'$
is a folding morphism iff 
\begin{itemize} 
  \item $\phi$ is total,
  \item $\forall t, t'\in T$. $(\pre{t} = \pre{t'}\ \land \phi_T(t) = \phi_T(t')) \Rightarrow t = t'$.
\end{itemize}
These requirements are standard for folding morphisms.
A folding what it does is to fold entirely a \mcn{} onto another (the requirement of totality of the
mapping) and it does in an economical way, as transitions that are not distinguishable in the target net
should be the same transition.

The algorithm will construct a spread net and also a morphism that will turn out to be a folding morphism.

\paragraph{Spreading algorithm:}\
We spread a \mcn{} with respect to a certain domain $\mathcal{A}$ of information inferred by the
net itself and a set of $\vec{\tau}$ of ticking mappings that obey to a schema 
(which basically states how conflicts are spread through the various components).
In fact, as it will become clear in the following, the schema for the ticking mappings can be seen as a
\emph{parameter} of the spreading, and it simply state how time is counted in each component, also in relation
with the other components.

\begin{figure}[!t]

\hrulefill 

\paragraph{Input:} A \mcn\ $\mathsf{N} = (\langle\place, T, F, m, \ell, \Sigma\rangle, \partition)$ of dimension 
$\ind{\mathsf{N}}$, a \vcd\ $\mathcal{A}$ of dimension $\dime{\mathcal{A}} = \ind{\mathsf{N}}$
such that for each $i\in \ind{\mathsf{N}}$. $\mathsf{alph}(\mathcal{A}_i) = \ell(T_i)$,
a set $\vec{\tau}$ of ticking mapping and a set of operations $\mathit{op}^k_J$ satisfying the
requirements of Definition~\ref{de:op}

\paragraph{Output:} At each step a spread net $\mathcal{O}$ and a folding mapping $\phi$ onto $\mathsf{N}$
\smallskip

\paragraph{Initialization step:} Create $\card{m}$ places for $\mathsf{O}$ and define a bijection 
$\phi_{\place} \colon m_O \to m$.
Define, for each $p\in m_O$, $h_O(p) = (\varepsilon, \dots, \varepsilon)$, and set $O = \langle m_O, \emptyset, \emptyset, m_O,\ell_O, \place\cup T\rangle$ with $\ell_O(p) = \phi_{\place}(p)$, obtaining the   
and \mcn\ $\mathsf{O} = (O,\partition_O)$ where $\partition_O(p) = \partition(\phi_{\place}(p))$.
Finally set $\mathcal{O} = (\mathsf{O},h_O)$. The $\phi$ mapping has just the component on places. 
Output $(\mathcal{O},\phi)$.

\paragraph{Recursion:} Consider the spread net constructed so far 
$\mathcal{O} = (\mathsf{O},h_O)$ and the mapping $\phi$.

Let $C$ be a configuration of $\mathsf{O} = (O,\partition_O)$, with 
$O = \langle\place_O, T_O, F_O, m_O, \ell_O, \place\cup T\rangle$, and consider 
$\hat{m} = \phi_{\place}(\mathit{mark}(C))$.
Let $t\in T$ be a transition such that $\pre{t} \subseteq \hat{m}$. 
Check if $T_0$ contains a transition $t'$ such that
$\pre{t'} \subseteq  \mathit{mark}(C)$ and $\phi_T(t') = t$. If yes consider another configuration, 
if not then
\begin{itemize}
 \item add $t'$ to $T_O$ and set $\phi'_T(t') = t$ and $\phi'_T(t'') = \phi_T(t'')$ for all $t''\in T_O$, 
 \item add to $F_O$ the set $F'_O = \setcomp{(p',t')}{p'\in \mathit{mark}(C)\ \land\ \phi_{\place}(p')\in \pre{t}}$, 
 \item for each $p \in \post{t}$, check if there is a place $p'\in\place_O$ such that 
     \begin{itemize}
       \item $\phi'_{\place}(p') = p$ and
       \item $h'(p') = \tau_{\partition_O(p')}(\mathit{op}^{\partition_O(p')}_{J}(\setcomp{h(p'')}{p''\in 
             \mathit{mark}(C)\ \land\ \phi_{\place}(p'')\in\pre{t}}),\ell'(t))$, where
             $J =\setcomp{\partition_O(p'')}{p''\in \mathit{mark}(C)\ \land\ \phi_{\place}(p'')\in\pre{t}}$.
     \end{itemize}         
       If yes, then simply add $(t',p')$ to $F'_O$. 
       If not then create a place $p'$, set
       $h'(p') = \tau_{\partition_O(p')}(\mathit{op}^{\partition_O(p')}_{J}(\setcomp{h(p'')}{p''\in 
             \mathit{mark}(C)\ \land\ \phi_{\place}(p'')\in\pre{t}}),\ell'(t))$, where
             $J =\setcomp{\partition_O(p'')}{p''\in \mathit{mark}(C)\ \land\ \phi_{\place}(p'')\in\pre{t}}$
       add it to $\place_O$. Add $(t',p')$ to $F'_O$ as well,
 \item extend $\phi_{\place}$ by setting $\phi'_{\place}(p') = p$, and
 \item set $\partition_O(p') = \phi_{\place}^{-1}(\partition_O(p))$      
\end{itemize}
Let $\place'$ the set of the new added places, let $O' = \langle\place_O\cup\place', T\cup\setenum{t'},
F_0\cup F'_O, m_O,\ell'_O,\place\cup T\rangle$ with $\ell'_O(x) = \ell_O(x)$ for $x\in\place_O\cup T_O$, 
$\ell'_O(t') = t$ and $\ell'_O(p) = \phi_{\place}(p)$ for each $p\in \place'$, and
$\partition'_O(p) = \partition(\phi_{\place}(p))$ for $p\in \place'$ and 
$\partition'_O(p) = \partition_O(p)$ for $p\in \place_O$. 

Output $\mathcal{O} = ((O',\partition'_O),h')$ and $\phi'$.

\hrulefill 

\caption{The spreading algorithm}\label{fig:spreading-alg}
\end{figure}

\begin{proposition}
 \label{prop:spreading-alg}
 Let $\mathsf{N} = (\langle\place, T, F, m, \ell, \Sigma\rangle, \partition)$ be a \mcn{} of dimension $\ind{\mathsf{N}}$
 such that $\ell\colon T\to\Sigma$ is total and injective.
 For each $i\in\setenum{1, \dots, \ind{\mathsf{N}}}$ let $\mathit{Eq}_i$ be a set of equations on $\ell(T_i)^{\ast}$, 
 where $T_i$ are the transitions of the $i$-th component of $\mathsf{N}$ and 
 $\mathcal{A}_i = \ell(T_{i})^{\ast}/\sim_{\mathit{Eq}_i}$.
 Let $\mathcal{A} = \times_{i=1}^{\ind{\mathsf{N}}}\mathcal{A}_i$ and let 
 $\vec{\tau} = \setcomp{\tau_i}{1\leq i\leq\ind{\mathsf{N}}}$ be a set of \emph{ticking} mapping with 
 $\tau_i : \mathcal{A}\times T_i \to \mathcal{A}$. Then the algorithm in Figure~\ref{fig:spreading-alg} produces
 a spreading net $\mathfrak{S}_{\vec{\tau}}^{\mathcal{A}}(\mathsf{N}) = (\mathsf{O}, h)$ 
 and a folding morphism $\phi \colon \mathsf{support}(\mathfrak{S}(\mathsf{N})) \to \mathsf{N}$.
\end{proposition}
It is quite obvious that the algorithm define a spread net and a folding morphism as well.
Observe that in the algorithm in Figure~\ref{fig:spreading-alg} we could have done the 
recursion step a bit differently, namely 
for each $t'$ added such that $\phi_T(t') = t$, we could have added $\card{\post{t}}$ places, and
then some of them could be
glued with some others in the spread net constructed so far provided that they are related to the same place in 
$\mathsf{N}$
and have the same vector-clock annotation. This alternative guarantees the fact that the morphism constructed is a folding one.

We want to stress that, depending on the vector-clock domain, the algorithm produce a \emph{finite} data structure 
(a finite spread net).
Indeed, if the elements of the vector-clock domain are finite the spread net constructed is finite as well, 
due to the way the labeling $\ell$ is defined when constructing the spread net. 

 \begin{figure}[!h]
 \centerline{
 \scalebox{1.1}{
 \begin{tikzpicture}
  [bend angle=45, scale=0.8, pre/.style={<-,shorten
    <=1pt,>=stealth',semithick}, post/.style={->,shorten
    >=1pt,>=stealth',semithick},place/.style={circle, draw=black,
    thick, minimum size = 5mm}, transition/.style={rectangle, draw=black, thick, minimum size = 5mm},
    posta/.style={-,semithick},
    invplace/.style={circle, draw=black!0,thick}]
\node[place, tokens = 1] (a) at (1,5) [label = above:\scriptsize{$(a,(\varepsilon,\varepsilon))$},label = below:\scriptsize{$p_1$}]{};
\node[place] (b) at (1,3)[label = below:\scriptsize{$(b,(s,\varepsilon))$},label = above:\scriptsize{$p_2$}]{};
\node[place] (c) at (2,1) [label = above:\scriptsize{$(c,(su,u))$},label = left:\scriptsize{$p_4$}]{};
\node[place, tokens = 1] (d) at (10,3) [label = above:\scriptsize{$(d,(\varepsilon,\varepsilon))$},,label = right:\scriptsize{$p_3$}]{};
\node[place] (e) at (10,1) [label = above:\scriptsize{$(e,(su,u))$},label = right:\scriptsize{$p_5$}]{};
\node[place] (a1) at (5.2,-1) [label = above:\scriptsize{$(a,(su,uz))\qquad \qquad$},label = below:\scriptsize{$\qquad p_6$}]{};
\node[place] (d1) at (8,-1) [label = above:\scriptsize{$\qquad (d,(su,uz))$},label = right:\scriptsize{$p_7$}]{};
\node[place] (b1) at (5.2,-3) [label = right:\scriptsize{$\quad (b,(s,uz))$},label = below:\scriptsize{$p_8$}]{};
\node[place] (b2) at (2,-6) [label = left:\scriptsize{$(b,(suv,u))$},label = right:\scriptsize{$p_9$}]{};
\node[place] (d2) at (11,-4.5) [label = right:\scriptsize{$(d,(su,uz))$},label = left:\scriptsize{$p_{10}$}]{};
\node[transition] (s1) at ( 0, 4 ) [label = left:\scriptsize{$t_1$}] {$s$}
edge[pre, bend left] (a)
edge[post, bend right] (b);
\node[transition] (t1) at ( 2, 4 ) [label = right:\scriptsize{$t_2$}] {$t$}
edge[pre, bend right] (a)
edge[post, bend left] (b);
\node[transition] (u1) at ( 6, 2 ) [label = above:\scriptsize{$t_3$}]{$u$}
edge[pre] (b)
edge[pre] (d)
edge[post] (c)
edge[post] (e);
\node[transition] (z1) at ( 6, 0 ) [label = above:\scriptsize{$t_4$}]{$z$}
edge[pre] (c)
edge[pre] (e)
edge[post] (a1)
edge[post] (d1);
\node[transition] (s2) at ( 4.2, -2 ) [label = left:\scriptsize{$t_5$}]{$s$}
edge[pre, bend left] (a1)
edge[post, bend right] (b1);
\node[transition] (t2) at ( 6.2, -2 ) [label = right:\scriptsize{$t_6$}]{$t$}
edge[pre, bend right] (a1)
edge[post, bend left] (b1);
\node[transition] (u1) at ( 6, -4.5 ) [label = below:\scriptsize{$t_7$}] {$u$}
edge[pre] (b1)
edge[pre, bend right] (d1)
edge[post, bend left] (c)
edge[post, bend right] (e);
\node[transition] (v) at (0,-2.5) [label = right:\scriptsize{$t_8$}] {$v$}
edge[pre] (c)
edge[post] (b2);
\node[transition] (w) at (11,-2.5) [label = left:\scriptsize{$t_9$}] {$w$}
edge[pre] (e)
edge[post] (d2);
\node[transition] (z1) at ( 3.6, -7.5 ) [label = below:\scriptsize{$t_{10}$}] {$u$}
edge[pre] (b2)
edge[pre, bend right] (d2)
edge[post] (a1)
edge[post, bend right] (e);
\end{tikzpicture}}
}

\caption{A spread net}\label{fig:spread-of-running-example}
\end{figure}
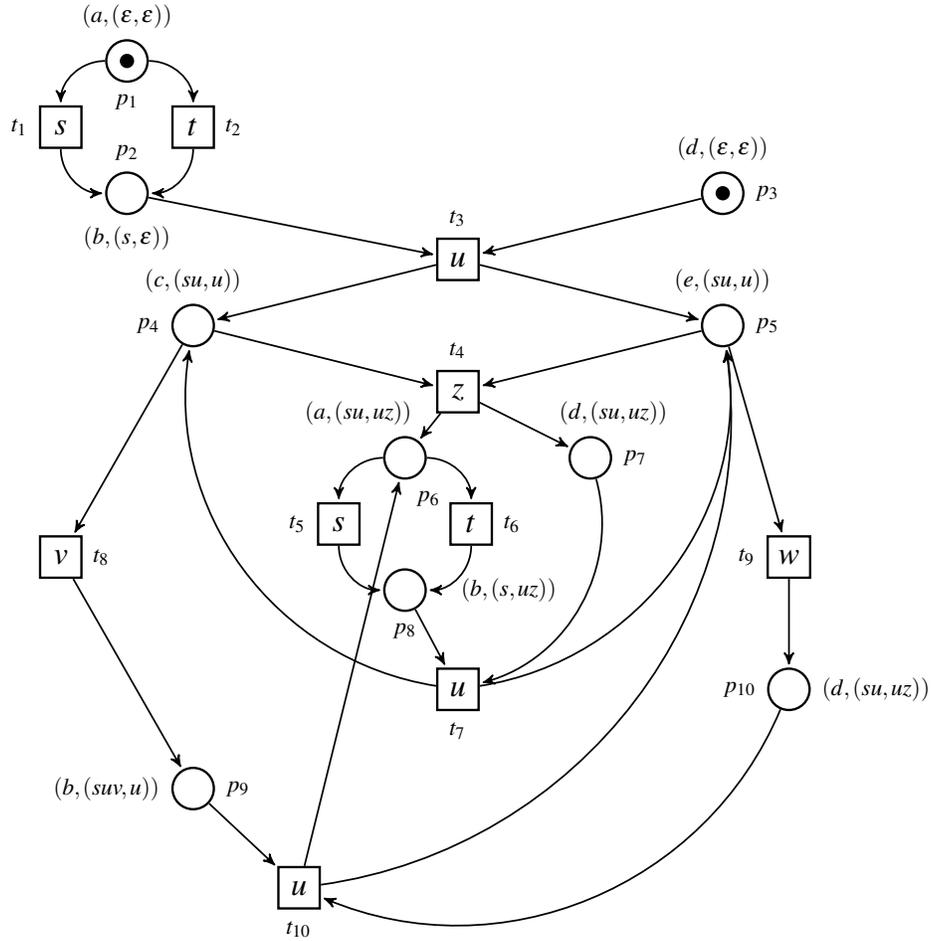

\begin{example}\label{ex:spread-example}
 Consider the net in Figure~\ref{fig:spread-of-running-example}. 
 This is the spreading of the \mcn{} in
 Figure~\ref{fig:mcn-running-example} according to the $\tau$s ticking mappings and
 to the vector-clock domain as described in the following.

 The ticking domain for the first component (the one on the left) is the with alphabet 
 $\mathsf{alph}(\mathcal{A}_1) = \setenum{s, t, u, v, z}$ and 
 the equations are $\varepsilon = \varepsilon$,  
 $s = t$, $su = tu$, $suz = su$, $tuz = su$, $sus = s$, $tus = s$, $sut = s$, $tut = s$, 
 $suvu = su$, $tuvu = su$, $u = \varepsilon$, $v = \varepsilon$, $ss = s$,
 $ts = s$ and $tt = t$ and $x = \varepsilon$ for each $x \in \mathsf{alph}(\mathcal{A}_1)^{\ast}$
 $\card{x}\geq 4$ and $x \neq suvu$ and $tuvu$. The  equivalence classes obtained are
 $\eqclass{\varepsilon}{\sim_1}$, $\eqclass{s}{\sim_1}$, $\eqclass{su}{\sim_1}$, $\eqclass{suv}{\sim_1}$ and 
 $\eqclass{suz}{\sim_1}$, and these form the ticking domain 
 $\mathcal{A}_1$. 
 Concerning the ticking domain for the second component, the alphabet is 
 $\mathsf{alph}(\mathcal{A}_2) = \setenum{u, w, z}$
 the equivalence relation is based on the following equations: $uwu = uz$, $\varepsilon = \varepsilon$ and
 for each other word $x$ in $\mathsf{alph}(\mathcal{A}_2)^{\ast}$ beside the ones involved
 in these equations, we have $x = \varepsilon$. 
 The equivalence classes
 we obtain are $\eqclass{\varepsilon}{\sim_2}, \eqclass{u}{\sim_2}$ and $\eqclass{uz}{\sim_2}$, 
 which are the elements of $\mathcal{A}_2$. 
 Equivalence classes are identified with their representative.
 
 The operations $\tau_i$ (with $i\in \setenum{1, 2}$) take a vector-clock and  concatenate
 each word with label $\ell(t)$, provided that $\ell(t)$ appears in the alphabet, thus
 $\tau_1((su,u),t_8) = (suv,u)$ as $\ell(t_8) = v$ is in the alphabet of the ticking domain $\mathcal{A}_1$
 but not in the alphabet of $\mathcal{A}_2$, and $\tau_1((su,u),t_8) = (suz,uz)$ as
 $\ell(t_{10}) = z$ belongs to both alphabets.
 
 The operations $\mathit{op}$ (we omit the indexes as it is clear what they do) if applied to just one
 vector return the same vector, otherwise the first component of the resulting vector comes from the first one
 and the second component from the second one.

 In the figure places are annotated with the pair $(p,\alpha)$ where $p$ is the name of the
 place in the net $\mathsf{N}$ in 
 Figure~\ref{fig:mcn-running-example} and $\alpha \in \mathcal{A}_1\times \mathcal{A}_2$.
 The first component of the pair is the $\ell$ mapping and the second is the $h$ mapping
 $h \colon \setenum{p_1, \dots, p_{10}} \to \mathcal{A}_1\times \mathcal{A}_2$ defined
 as follows; $h(p_1) = (\varepsilon,\varepsilon) = h(p_3), h(p_2) = (s,\varepsilon), 
 h(p_4)  = h(p_5)  = (su,u)$, $h(p_6) = (suz,uz) = h(p_7), h(p_8) = (s,uz), h(p_9) = (suv,u)$ and $h(p_{10}) = (su,uw)$.
 
 \end{example}

Observe that the spread net is finite as the \vcd\ is finite and  $\ell$ maps places and transitions of the
spread net onto a finite set. 
Another spread net over the same domain with the same ticking mappings does not need to be finite, provided that the
$\ell$ mapping has an infinite codomain.

The nice property that the spreading of a net enjoys  is that it is indeed a universal construction.

\begin{theorem}
 \label{th:spreding-is-universal}
 Let $\mathsf{N}$ be an \mcn,  
 then for each $\mathcal{S}$ spread net with respect to $\vec{\tau}$ and $\mathcal{A}$
 and morphism $g \colon \mathsf{support}(\mathcal{S}) \to \mathsf{N}$,
 there exists a unique morphism 
 $l \colon \mathcal{S} \to \mathfrak{S}_{\vec{\tau}}^{\mathcal{A}}(\mathsf{N})$ such that 
 $g = \mathfrak{F}(l) \circ \phi$.
\end{theorem}
The theorem implies that the spreading of \mcn\ with respect to a 
given $\vec{\tau}$ and $\mathcal{A}$, is somehow the \emph{best} construction with these characteristic we
can aim at. 
The fact that it is the best construction depends on the way the spreading is performed not only for the
annotation of places but also for the labeling of them. It is then quite obvious that any other spread net which is
mapped onto the one to be spread, should have a different labeling has it cannot have a different annotation.

We substantiate our claim showing that this new notion covers various notion of unfoldings. 
We will consider here just branching processes and trellis processes (for the proper definitions we refer to
\cite{Eng:BPPN,Win:ES} for branching processes and \cite{Fab:Trellis} for trellis processes).

\paragraph{Branching Processes:}
The ticking domain to be considered in this case is, for each component, the one induced by the set of 
equations containing just $\varepsilon = \varepsilon$, and the alphabet of each
ticking domain are the transitions of the component. The result is that each
equivalence class contains just a word. 
We call the resulting vector-clock domain $\mathcal{A}_{BP}$.
The $\tau_i$ add the transition to the words in the entries of the vector-clock that are
involved in the synchronization, and we call these $\tau_i$ as $\vec{\tau}_{BP}$. 
The operations $\mathit{op}$ take the words belonging to the components synchronizing, shuffle them taking
into account the synchronization transitions, and produces a new
vector-clock where each entry is the word obtained projecting on the proper alphabet the word obtained as 
described above.
This convey the intuition that from a given transition, there is a unique path to the places in the initial
marking, which is the one of a causal net (\cite{Win:ES}).

We can state the following result, where, $\mathcal{U}_{BP}(\mathsf{N})$ is the branching process obtained by the \mcn{} $\mathsf{N}$.

 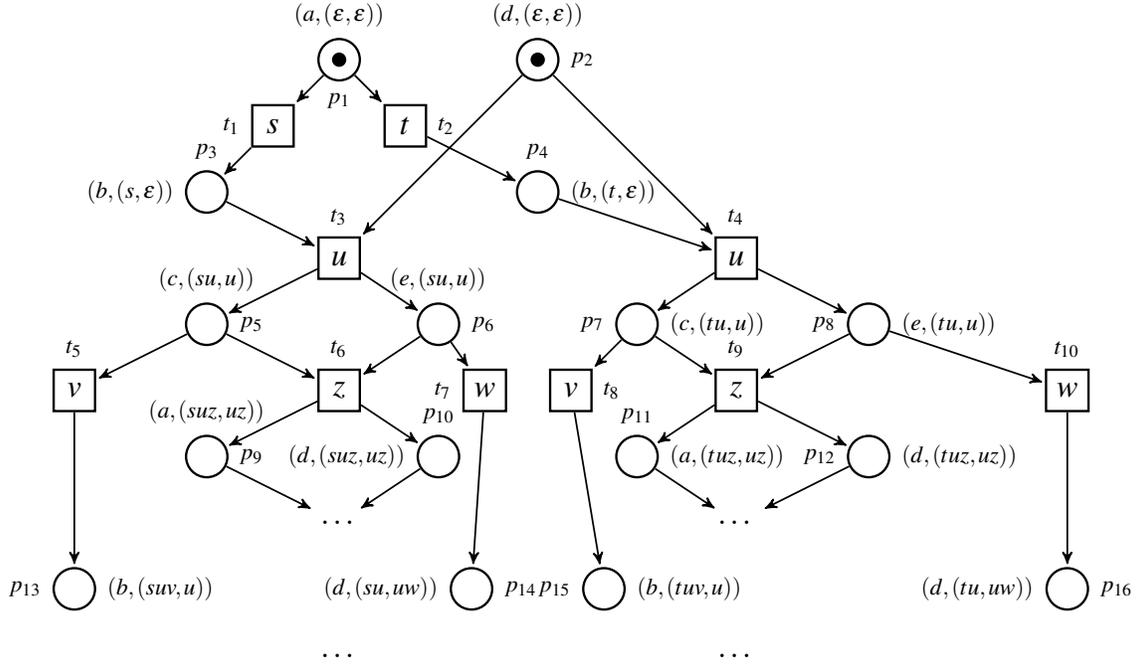
\begin{figure}[!t]
 \centerline{
 \scalebox{1.1}{
 \begin{tikzpicture}
  [bend angle=45, scale=0.8, pre/.style={<-,shorten
    <=1pt,>=stealth',semithick}, post/.style={->,shorten
    >=1pt,>=stealth',semithick},place/.style={circle, draw=black,
    thick, minimum size = 5mm}, transition/.style={rectangle, draw=black, thick, minimum size = 5mm},
    posta/.style={-,semithick},
    invplace/.style={circle, draw=black!0,thick}]
\node[place, tokens = 1] (a) at (3,5) [label = above:\scriptsize{$(a,(\varepsilon,\varepsilon))$},label = below:\scriptsize{$p_1$}]{};
\node[place, tokens = 1] (d) at (6,5) [label = above:\scriptsize{$(d,(\varepsilon,\varepsilon))$},,label = right:\scriptsize{$p_2$}]{};
\node[place] (b) at (1,3)[label = left:\scriptsize{$(b,(s,\varepsilon))$},label = above:\scriptsize{$p_3$}]{};
\node[place] (bb) at (6,3)[label = right:\scriptsize{$(b,(t,\varepsilon))$},label = above:\scriptsize{$p_4$}]{};
\node[place] (c) at (1,1) [label = above:\scriptsize{$(c,(su,u))$},label = right:\scriptsize{$p_5$}]{};
\node[place] (cc) at (7.5,1) [label = right:\scriptsize{$(c,(tu,u))$},label = left:\scriptsize{$p_7$}]{};
\node[place] (e) at (4.5,1) [label = above:\scriptsize{$(e,(su,u))$},label = right:\scriptsize{$p_6$}]{};
\node[place] (ee) at (11,1) [label = right:\scriptsize{$(e,(tu,u))$},label = left:\scriptsize{$p_8$}]{};
\node[place] (a1) at (1,-1) [label = above:\scriptsize{$(a,(suz,uz))$},label = right:\scriptsize{$p_9$}]{};
\node[place] (d1) at (4.5,-1) [label = left:\scriptsize{$\qquad (d,(suz,uz))$},label = above:\scriptsize{$p_{10}$}]{};
\node[place] (aa1) at (7.5,-1) [label = right:\scriptsize{$(a,(tuz,uz))$},label = above:\scriptsize{$p_{11}$}]{};
\node[place] (dd1) at (11,-1) [label = right:\scriptsize{$(d,(tuz,uz))$},label = left:\scriptsize{$p_{12}$}]{};
\node[place] (b1) at (-1,-3) [label = right:\scriptsize{$(b,(suv,u))$},label = left:\scriptsize{$p_{13}$}]{};
\node[place] (d2) at (5,-3) [label = left:\scriptsize{$(d,(su,uw))$},,label = right:\scriptsize{$p_{14}$}]{};
\node[place] (b2) at (7,-3) [label = right:\scriptsize{$(b,(tuv,u))$},label = left:\scriptsize{$p_{15}$}]{};
\node[place] (d3) at (14,-3) [label = left:\scriptsize{$(d,(tu,uw))$},,label = right:\scriptsize{$p_{16}$}]{};
\node[transition] (s1) at ( 2, 4 ) [label = left:\scriptsize{$t_1$}] {$s$}
edge[pre] (a)
edge[post] (b);
\node[transition] (t1) at ( 4, 4 ) [label = right:\scriptsize{$t_2$}] {$t$}
edge[pre] (a)
edge[post] (bb);
\node[transition] (u1) at ( 3, 2 ) [label = above:\scriptsize{$t_3$}]{$u$}
edge[pre] (b)
edge[pre] (d)
edge[post] (c)
edge[post] (e);
\node[transition] (uu1) at ( 9, 2 ) [label = above:\scriptsize{$t_4$}]{$u$}
edge[pre] (bb)
edge[pre] (d)
edge[post] (cc)
edge[post] (ee);
\node[transition] (z1) at ( 3, 0 ) [label = above:\scriptsize{$t_6$}]{$z$}
edge[pre] (c)
edge[pre] (e)
edge[post] (a1)
edge[post] (d1);
\node[transition] (zz1) at ( 9, 0 ) [label = above:\scriptsize{$t_9$}]{$z$}
edge[pre] (cc)
edge[pre] (ee)
edge[post] (aa1)
edge[post] (dd1);
\node[transition] (v) at (-1,0) [label = above:\scriptsize{$t_5$}] {$v$}
edge[pre] (c)
edge[post] (b1);
\node[transition] (w) at (5.2,0) [label = left:\scriptsize{$t_7$}] {$w$}
edge[pre] (e)
edge[post] (d2);
\node[transition] (ww) at (14,0) [label = above:\scriptsize{$t_{10}$}] {$w$}
edge[pre] (ee)
edge[post] (d3);
\node[transition] (vv) at (6.5,0) [label = right:\scriptsize{$t_8$}] {$v$}
edge[pre] (cc)
edge[post] (b2);
\node (1) at (3, -2) {$\dots$}
edge[pre] (a1)
edge[pre] (d1);
\node (2) at (9, -2) {$\dots$}
edge[pre] (aa1)
edge[pre] (dd1);
\node (1) at (3, -4) {$\dots$};
\node (2) at (9, -4) {$\dots$};
\end{tikzpicture}}
}
\caption{The initial part of the Branching process of the \mcn{} $\mathcal{N}$ in Figure~\ref{fig:mcn-running-example}}\label{fig:spreadBP-of-running-example}
\end{figure}

\begin{proposition}
 \label{pr:BP-Spread}
 Let $\mathsf{N}$ be a \mcn{}, and let $\mathfrak{S}_{\vec{\tau}_{BP}}^{\mathcal{A}_{BP}}(\mathsf{N})$ 
 be its spreading with respect to $\vec{\tau}_{BP}$ and $\mathcal{A}_{BP}$. 
 
 Then
 $\mathsf{support}(\mathfrak{S}_{\vec{\tau}_{BP}}^{\mathcal{A}_{BP}}(\mathsf{N}))$ is isomorphic to 
 $\mathcal{U}_{BP}(\mathsf{N})$.
\end{proposition}

\begin{example}\label{ex:BP-spread} 
 Figure~\ref{fig:spreadBP-of-running-example} shows just the first part of the spreading of $\mathsf{N}$ 
 according to the
 $\mathcal{A}_{BP}$ domain. The object constructed in this way is, as expected, infinite. 
 Conflicts are inherited along the causality
 paths (executions in each automata) and the \emph{quantity} of information associated to 
 each place in this spreading increases. 
 The annotation of a place contains, for each components, the trace in this component leading to that 
 place.
 For instance, consider the place $p_9$. The annotation of $p_9$ is $(suz,uz)$. 
 In fact the two components of the net synchronize first 
 on $u$ and then on $z$, in the first component the first transition executed is $s$ whereas the second component
 should synchronize. Hence the annotation regarding the first component is $suz$ and the one regarding the second
 component is $uz$.
\end{example}

\paragraph{Trellises:}
Here the ticking domain for each component 
is the one induced by the following set of equations: $u = v$ for all $u, v$ words with the same
length on the alphabet such that they correspond to a firing sequence in the component ending in the same
place (as the length counts there is a difference with the words ending in the same place). Thus two
words $w$ and $w'$ are equivalent iff they have the same length and
if they put a token in the same place. We call this domain $\mathcal{A}_{Tr}$.

The $\tau_i$ work as follows: each of
them receives in input a vector-clock and a transition and return a vector-clock where the transition 
is concatenated to the word in the proper entry, and
all the others are set to $\varepsilon$. The set of these $\tau_i$ is called 
$\vec{\tau}_{Tr}$. 
The operations $\mathit{op}$ work like the ones devised for the branching processes.

With $\mathcal{U}_{Tr}(\mathsf{N})$ we denote the trellis obtained by the \mcn{} $\mathsf{N}$ 
we have the following 
result:
\begin{proposition}
 Let $\mathsf{N}$ be a \mcn{}, and let $\mathfrak{S}_{\vec{\tau}_{Tr}}^{\mathcal{A}_{Tr}}(\mathsf{N})$ 
 be its spreading 
 with respect to $\vec{\tau}_{Tr}$ and $\mathcal{A}_{Tr}$. 
 
 Then
 $\mathsf{support}(\mathfrak{S}_{\vec{\tau}_{Tr}}^{\mathcal{A}_{Tr}}(\mathsf{N}))$ is isomorphic to  
 $\mathcal{U}_{Tr}(\mathsf{N})$.
\end{proposition}
\begin{example}\label{ex:Tr-spread}
 Figure~\ref{fig:trellis-of-running-example} shows the first part of the spreading of $\mathsf{N}$ according to the
 $\mathcal{A}_{Tr}$ domain. The object constructed in this way is again infinite. Conflicts are folded in each 
 automata according to the length of the local executions. 
  
  \begin{figure}[!th]
 \centerline{
 \scalebox{1.1}{
 \begin{tikzpicture}
  [bend angle=45, scale=0.8, pre/.style={<-,shorten
    <=1pt,>=stealth',semithick}, post/.style={->,shorten
    >=1pt,>=stealth',semithick},place/.style={circle, draw=black,
    thick, minimum size = 5mm}, transition/.style={rectangle, draw=black, thick, minimum size = 5mm},
    posta/.style={-,semithick},
    invplace/.style={circle, draw=black!0,thick}]
\node[place, tokens = 1] (a) at (1,5) [label = above:\scriptsize{$(a,(\varepsilon,\varepsilon))$},label = below:\scriptsize{$p_1$}]{};
\node[place] (b) at (1,3)[label = below:\scriptsize{$(b,(s,\varepsilon))$},label = above:\scriptsize{$p_2$}]{};
\node[place, tokens = 1] (d) at (9,3) [label = above:\scriptsize{$(d,(\varepsilon,\varepsilon))$},,label = right:\scriptsize{$p_3$}]{};
\node[place] (c) at (2,1) [label = above:\scriptsize{$(c,(su,\varepsilon))$},label = left:\scriptsize{$p_4$}]{};
\node[place] (e) at (9,1) [label = above:\scriptsize{$(e,\varepsilon,u))$},label = right:\scriptsize{$p_5$}]{};
\node[place] (a1) at (4,-1) [label = above:\scriptsize{$(a,(suz,\varepsilon))\qquad \qquad$},label = below:\scriptsize{$p_6$}]{};
\node[place] (d1) at (8,-2.5) [label = right:\scriptsize{$(d,(\varepsilon,uz))$},label = left:\scriptsize{$p_7$}]{};
\node[place] (b1) at (4,-3) [label = left:\scriptsize{$(b,(suzs,\varepsilon))\quad $},label = above:\scriptsize{$p_8$}]{};
\node[place] (b2) at (0,-3) [label = left:\scriptsize{$(b,(suv,\varepsilon))$},label = right:\scriptsize{$p_9$}]{};
\node[place] (e1) at (8,-6) [label = right:\scriptsize{$(e,(\varepsilon,uwu))$},label = left:\scriptsize{$p_{11}$}]{};
\node[place] (c2) at (5,-6) [label = left:\scriptsize{$(c,(suzsu,\varepsilon))$},label = above:\scriptsize{$p_{10}$}]{};
\node[place] (c3) at (1,-6) [label = left:\scriptsize{$(c,(suvu,\varepsilon))$},label = right:\scriptsize{$p_{12}$}]{};
\node[transition] (s1) at ( 0, 4 ) [label = left:\scriptsize{$t_1$}] {$s$}
edge[pre, bend left] (a)
edge[post, bend right] (b);
\node[transition] (t1) at ( 2, 4 ) [label = right:\scriptsize{$t_2$}] {$t$}
edge[pre, bend right] (a)
edge[post, bend left] (b);
\node[transition] (u1) at ( 6, 2 ) [label = above:\scriptsize{$t_3$}]{$u$}
edge[pre] (b)
edge[pre] (d)
edge[post] (c)
edge[post] (e);
\node[transition] (z1) at ( 6, 0 ) [label = above:\scriptsize{$t_4$}]{$z$}
edge[pre] (c)
edge[pre] (e)
edge[post] (a1)
edge[post] (d1);
\node[transition] (s2) at ( 3, -2 ) [label = left:\scriptsize{$t_5$}]{$s$}
edge[pre, bend left] (a1)
edge[post, bend right] (b1);
\node[transition] (t2) at ( 5, -2 ) [label = right:\scriptsize{$t_6$}]{$t$}
edge[pre, bend right] (a1)
edge[post, bend left] (b1);
\node[transition] (u1) at ( 8, -4.5 ) [label = right:\scriptsize{$t_7$}] {$u$}
edge[pre] (b1)
edge[pre] (d1)
edge[post] (c2)
edge[post] (e1);
\node[transition] (uu1) at ( 4.4, -4.5 ) [label = above:\scriptsize{$t_8$}] {$u$}
edge[pre] (b2)
edge[pre] (d1)
edge[post] (c3)
edge[post] (e1);
\node[transition] (v) at (0,-1.5) [label = right:\scriptsize{$t_9$}] {$v$}
edge[pre] (c)
edge[post] (b2);
\node[transition] (w) at (9,-1) [label = left:\scriptsize{$t_{10}$}] {$w$}
edge[pre] (e)
edge[post] (d1);
\node (1) at ( 1, -7) {$\vdots$};
\node (2) at ( 5, -7) {$\vdots$};
\node (2) at ( 8, -7) {$\vdots$};
\end{tikzpicture}}
}

\caption{The initial part of the trellis of the \mcn{} $\mathcal{N}$ in Figure~\ref{fig:mcn-running-example}}\label{fig:trellis-of-running-example}
\end{figure}
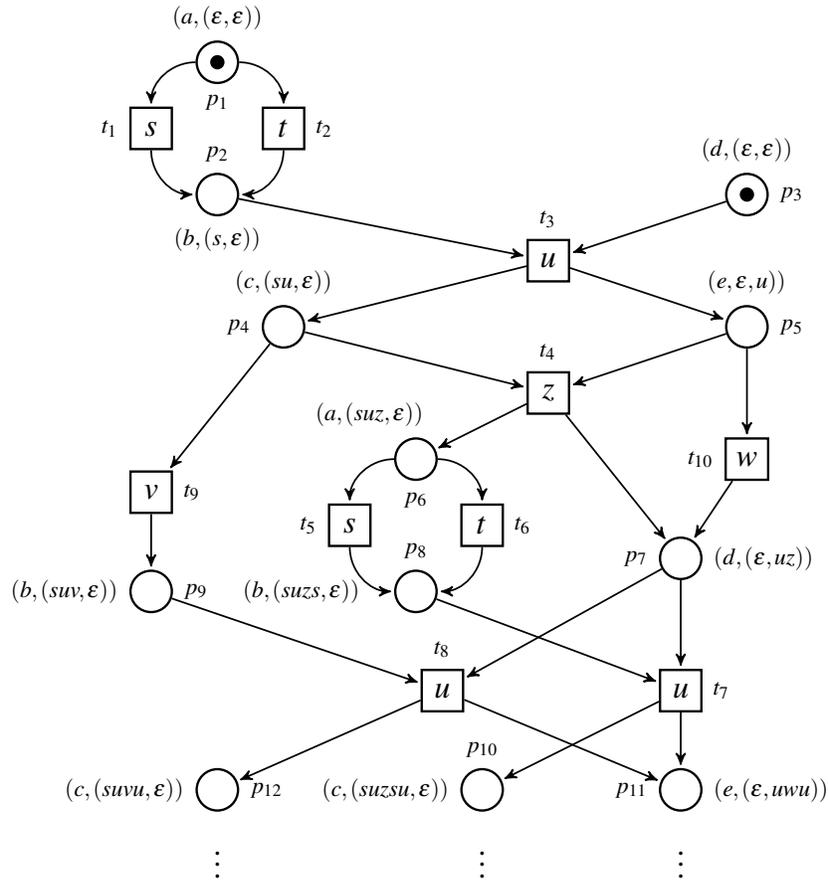
 
 Each place is annotated with the language of all the traces leading to the place (with respect to the
 equivalence relation).
 Consider the place $p_{12}$. It belongs to the first component, and the \emph{distance} from $p_1$ is $4$.
 The two words of belonging to the language of the first automaton ending in the image of $p_{12}$ ($c$) are 
 $suvu$ and $tuvu$. 
 If we consider the place $p_{11}$, it belongs to the second component, and the annotation $uwu$ is the equivalence 
 class containing also $uzu$, which are the two words of length $3$ ending in the image of $p_{11}$ ($e$).
\end{example}

\section{Conclusions}
In this paper we have presented the notion of spread net which is able to
represent the non sequential behaviors of safe nets, in particular of \mcn{s}.
A spread net is a net where each place has an annotation representing the amount of information
that has been collected to \emph{produce} that place, and the information depends on two elements. One
element is the information inferred from the annotations of the places
in the preset of the transitions in the preset of that place, and the second element is the transition
itself. 

Beside the notion of spread net we have formalized the algorithm for spreading a net, which is basically
the same algorithm which is used to unfold a net. Here we have presented the usual one based on the
notion of configuration of a spread net, but the annotations of places may be used to define 
more easily which subset of the involved places is a part of a marking reachable in the spread net and
henceforth corresponding to a marking of the unfolded net.

Here we have considered very simple domains, without making any real consideration on the kind of
properties one would like to prove on spread nets. But the main advantage of the notion is the fact
that it is indeed independent on the chosen domain, hence it can be used in quite different context.

In this paper we have not investigated an interesting issue, namely what is the brand of event structure 
related to spread net, like it is done in \cite{GP:CSESPN}. However we believe that configuration structures
can be easily related with spread nets, hence part of the results presented there should be applicable
also in our setting. 
Clearly the kind of event structure related to spread nets will be somehow parametric on the kind of annotations
of the spread net.

{\small \paragraph{Acknowledgments.}
This work is partially supported by Aut. Reg. of Sardinia
projects ``Sardcoin'' and ``Smart collaborative engineering''. The authors wish to thank the ICEcreamers and the anonymous reviewers for their useful comments, suggestions and criticisms.

\bibliography{biblio}

\end{document}